\documentclass[aps,prd,nofootinbib,showpacs,preprintnumbers]{revtex4-1}
\usepackage[usenames]{color}
\usepackage[dvips]{epsfig}
\usepackage{amssymb}
\usepackage{verbatim}
\usepackage{psfrag}
\newcommand{\beqn} {\begin{equation}}
\newcommand{\eqn} {\end{equation}}
\newcommand{\hm}{\hat{m}}

\def \hmu{\hat{\mu}}
\def \beq{\begin{equation}}
\def \eeq{\end{equation}}
\def \bea{\begin{eqnarray}}
\def \eea{\end{eqnarray}}

\def \bet0{\beta_0}
\def \bet1{\beta_1}
\def \simgt{\,\rlap{\lower 7.5 pt\hbox{$\mathchar \sim$}}\raise 3 pt \hbox{$>$}\,}
\def \simlt{\,\rlap{\lower 7.5 pt\hbox{$\mathchar \sim$}}\raise 3 pt \hbox{$<$}\,}

\def\lsim{\raise0.3ex\hbox{$<$\kern-0.75em\raise-1.1ex\hbox{$\sim$}}}
\def\gsim{\raise0.3ex\hbox{$>$\kern-0.75em\raise-1.1ex\hbox{$\sim$}}}

\begin{document}
\title{The phase boundary for the chiral transition in (2+1)-flavor QCD \\[2mm]
at small values of the chemical potential}
\author{O. Kaczmarek$^{\rm a}$,
F. Karsch$^{\rm a,b}$, E. Laermann$^{\rm a}$, C. Miao$^{\rm b}$, S. Mukherjee$^{\rm b}$,\\
P. Petreczky$^{\rm b}$, 
C. Schmidt$^{\rm c,d}$, W. Soeldner$^{\rm c,d}$ and W. Unger$^{\rm a}$
}
\affiliation{
$^{\rm a}$ Fakult\"at f\"ur Physik, Universit\"at Bielefeld, D-33615 Bielefeld, Germany\\
$^{\rm b}$ Physics Department, Brookhaven National Laboratory,Upton, NY 11973, USA \\
$^{\rm c}$ Frankfurt Institute for Advanced Studies, J.W.Goethe 
Universit\"at Frankfurt, \\
D-60438 Frankfurt am Main, Germany\\
$^{\rm d}$ GSI Helmholtzzentrum f\"ur
Schwerionenforschung, Planckstr.~1, D-64291 Darmstadt, Germany
}

\begin{abstract}
We determine the chiral phase transition line in
(2+1)-flavor QCD for small values of the light quark 
chemical potential.
We show that for small values of the chemical potential the curvature of 
the phase transition line can be deduced from an analysis of scaling 
properties of the chiral condensate and its susceptibilities.
To do so we extend earlier studies of the magnetic equation of state
in (2+1)-flavor QCD to finer lattice spacings, $aT=1/8$. We use
these universal scaling properties of 
the chiral order parameter to extract the curvature of the 
transition line at two values of the cut-off, $aT=1/4$ and $1/8$.
We find that cut-off effects are small for the curvature parameter 
and determine the transition line in the chiral limit 
to leading order in the light quark chemical potential. 
We obtain
$T_c(\mu_q)/T_c(0) = 1 - 0.059(2)(4) (\mu_q/T)^2 +{\cal O}(\mu_q^4)$.
\end{abstract}
\preprint{BI-TP 2010/42}
\maketitle

\section{Introduction}

Extending lattice QCD calculations to non-zero baryon-chemical potential or,
equivalently, to non-zero net baryon number density is known to be difficult
in general. However, important information on the QCD phase diagram
can be deduced for small values of the chemical potential
by using well established numerical techniques such as reweighting \cite{Fodor},
analytic continuation \cite{analytic,Lombardo} or Taylor expansion 
\cite{Taylor,Gavai}.
We will concentrate here on the latter approach. 

Not only do Taylor expansions
of thermodynamic observables provide information on QCD thermodynamics at
small but non-zero chemical potential, the expansion coefficients themselves
also are sensitive indicators for critical behavior in the vicinity
of the chiral phase transition at vanishing chemical potential. 
As the chemical potential couples to the quark number current, which does
not break chiral symmetry, it acts to leading order 
like a temperature variable. Derivatives with respect to chemical potentials
lead to susceptibilities which exhibit critical behavior similar to that
of thermal susceptibilities \cite{mucrit}. We will show here that a 
calculation of the leading order Taylor expansion coefficient of the chiral 
order parameter, which defines a mixed susceptibility, allows to perform 
quantitative studies of the phase boundary between low and high temperature 
phases of QCD close to $\mu=0$.

At non-zero values of the chemical potential a phase boundary in the 
temperature and chemical potential parameter space of QCD is well defined 
only in the heavy quark limit or for vanishing quark masses. In the former 
case the phase transition line corresponds to the first order deconfinement 
transition in the pure gauge  theory. At infinite values of the quark mass 
this transition is independent of the chemical potential and defines a 
straight line in the $T$-$\mu$ plane.
For a large range of quark mass values the transition line is not unique.  
It characterizes a region of (rapid) crossover in thermodynamic quantities 
and a pseudo-critical temperature extracted from these observables  
may differ somewhat, depending on the observable that is used.
In the chiral limit, however, the transition line is again well defined. 
For sufficiently large strange quark mass it 
defines a line of second order phase transitions in the universality class
of three dimensional $O(4)$ symmetric spin models \cite{Pisarski}. 

Taylor expansions, analytic continuation as well as reweighting techniques
have been used to locate the crossover line $T_c(\mu)$ in the 
$T$-$\mu$ plane for small values of the chemical potential \cite{Philipsen}. 
These calculations, which have been performed for different flavors
and various values of the quark masses, suggest that 
the curvature is small, {\it i.e.} $T_c(\mu)/T_c(0)$ decreases only by a 
few percent at $\mu/T\simeq 1$. 
However, so far most 
lattice QCD calculations performed to determine the transition line have been 
performed on coarse lattices\footnote{An attempt to determine the transition line closer
to the continuum limit has been presented in \cite{Endrodi}.}, {\it i.e.} 
lattices with only four sites in the temporal direction for which the lattice 
spacing in units of the temperature thus equals $aT=1/4$. 
Better control over the extrapolation to the continuum limit and the
quark mass dependence of the transition line clearly is needed.

So far studies of the transition line concentrated on its dependence
on a single chemical potential, taken to be either identical for the light
up ($\mu_u$) and down ($\mu_d$) quarks or of opposite sign. The 
former is the light quark chemical potential, $\mu_q=(\mu_u+\mu_d)/2$ and
latter is the isospin chemical potential, $\mu_I =(\mu_u-\mu_d)/2$ for which 
direct numerical calculations are possible.
In order to make contact to the situation met in heavy ion collisions
\cite{andro1,cleymans} one eventually wants to  analyze the influence of 
non-zero charge ($\mu_Q$) and strangeness ($\mu_S$) chemical potentials on 
the curvature of the transition line, {\it i.e.}, one should allow for non-
vanishing up-, down- and strange-quark chemical potentials.
For small values of the chemical potential this is possible in the 
framework we will outline here. At present we will, however,
restrict our discussion to the case of 
vanishing strange quark and isospin (electric charge) chemical potentials. 

We concentrate on an analysis of the phase transition
line in the chiral limit ($m_u=m_d=0$, $m_s>0$) where its dependence 
on $\mu_q$ is expected to be largest.
We will present a calculation of the critical line for small values of 
the light quark masses in the scaling regime of the finite temperature 
chiral phase transition. This allows to use scaling relations to 
extract the curvature of the phase transition line in the
chiral limit of QCD. The scaling relations naturally relate the
curvature of the critical line as function of $\mu_q$  
to the magnitude of a mixed susceptibility. 

We will perform our numerical calculations for
(2+1)-flavor QCD keeping the heavier strange quark mass close to its 
physical value and decreasing the two degenerate light quark masses towards 
the massless limit. On coarse lattice with temporal extent $N_\tau =4$
we will make use of a recently performed scaling analysis \cite{magnetic}
of the chiral order parameter performed with an improved staggered fermion
action.
This study showed that the chiral order parameter is well described by
a universal scaling function characteristic for a three dimensional, $O(N)$ 
universality class. As we are using a staggered fermion discretization
scheme for our scaling analysis we expect that the transition in the chiral 
limit, performed at non-zero lattice spacing, is controlled by the $O(2)$ 
rather than $O(4)$ universality class. We thus will analyze our numerical 
results in terms of $O(2)$ scaling functions. We will comment on the 
application of $O(4)$ scaling relations later on.
 
This paper is organized as follows. In the next section we will extend
the scaling analysis of the chiral order parameter to lattices
with temporal extent $N_\tau=8$. This provides the basic parameters 
needed for a calculation of the curvature of the chiral phase transition
line which will be discussed in Section III. We give a discussion of
our results and an outlook in Section IV.
  
\section{Magnetic equation of state}

In the vicinity of a critical point regular contributions to the partition
functions become negligible in higher order derivatives and the singular 
behavior of response functions will generally be dominated by contributions 
arising from the singular part of the free energy density\footnote{For systems
belonging to the 3-dimensional $O(2)$ or $O(4)$ universality classes this 
does not hold for the thermal response function (specific heat) as the 
relevant critical exponent $\alpha$ is negative in these cases.}
\begin{equation}
f(T,m_l,m_s,\mu_q,\mu_s) = f_s(T,m_l,m_s,\mu_q,\mu_s) + 
f_r(T,m_l,m_s,\mu_q, \mu_s) \; .
\label{freeenergy}
\end{equation} 
In addition to the temperature $T$, light ($m_l)$ and strange ($m_s$) quark 
masses we also allow for a dependence of the free energy density on the quark 
chemical potentials.
Close to the chiral phase transition temperature at vanishing chemical 
potential the singular part $f_s$ will give rise to universal scaling 
properties of response functions. This has been exploited to analyze 
basic universal features of the QCD phase diagram close to criticality
\cite{Hatta}.
 
The singular part of the free energy density
depends on the parameters of the QCD Lagrangian, e.g. the quark masses, and 
the external control parameters, temperature  and chemical potentials, 
only through two relevant couplings. These scaling variables, $t$ and $h$, 
control deviations from criticality, $(t,h)=(0,0)$, along
the two relevant directions, which in the case of QCD characterize 
fluctuations of the energy and chiral condensate, respectively. 
To leading order the scaling variable $h$ depends only on parameters that 
break chiral symmetry in the light quark sector, while $t$ depends on all 
other couplings. In particular, $t$ will depend on the light quark chemical 
potential
while $h$ remains unaffected by these in leading order,
\begin{eqnarray}
t &\equiv& \frac{1}{t_0}\left( \frac{T-T_c}{T_c} + 
\kappa_q \left( \frac{\mu_q}{T}\right)^2 
\right)
\ ,
\nonumber \\
h &\equiv& \frac{1}{h_0} \frac{m_l}{m_s} \ ,
\label{scalingfields}
\end{eqnarray}  
where $T_c$ is the phase transition temperature in the chiral limit and
$t_0$, $h_0$ are non-universal scale parameters (as is $T_c$).
While the combination $z_0=h_0^{1/\beta\delta}/t_0$ is
unique for a given theory, the values of $t_0$ and $h_0$ will change under
rescaling of the order parameter \cite{magnetic}. Note also that they depend on
the definition of the parameter introduced to control symmetry breaking, 
{\it i.e.} the fact that we choose the strange quark mass to normalize the 
symmetry breaking light quark mass parameter.
For the rest of this chapter we will not need to refer any further to the
contribution of chemical potentials to the reduced temperature $t$. We will
come back to it in the next chapter.

The singular part of the free energy, $f_s$, is a homogeneous function of 
its arguments. This can be used to rewrite it in terms of
the scaling variable $z=t/h^{1/\beta\delta}$ as
\begin{equation}
f_s(t,h) = h^{1+1/\delta} f_s(z,1) \equiv h^{1+1/\delta} f_s(z) \ .
\label{fs}
\end{equation}
where $\beta,\ \delta$ are critical exponents of the three dimensional $O(N)$ 
universality class \cite{Engels2001,Engels2003}, $\beta=0.349$ and 
$\delta=4.780$ for three dimensional
$O(2)$ models and $\beta=0.380$ and $\delta=4.824$ for $O(4)$, respectively. 
All parameters entering the definition
of $t$ and $h$, {\it i.e.} $t_0$, $h_0$ and $T_c$ may depend on the strange 
quark mass, but are otherwise unique in the continuum limit of (2+1)-flavor 
QCD. Just like the transition temperature $T_c$, however, also $t_0$ and $h_0$
are cut-off dependent and will need to be extrapolated to the continuum limit. 

The universal critical behavior of the order parameter, 
$M\sim \partial f/\partial m_l$, 
is controlled by a scaling function $f_G$ that arises from
the singular part of the free energy density after taking a derivative
with respect to the light quark mass,
\begin{equation}
M(t,h) \;=\; h^{1/\delta} f_G(z) \; .
\label{order}
\end{equation}
The scaling function $f_G(z)$ is well-known for the $O(2)$ and $O(4)$ 
universality classes through studies of three dimensional spin models 
\cite{engelsO2}.
This so-called magnetic equation of state, Eq.~\ref{order}, has been analyzed 
recently for (2+1)-flavor QCD using an improved staggered fermion formulation 
(p4-action) on lattices with temporal extent 
$N_\tau=4$ \cite{magnetic} and light quark masses as small as $m_l/m_s=1/80$, 
which corresponds to a pion mass that is about half its physical value.
It could be shown 
that the chiral order parameter can be mapped onto a universal $O(N)$ 
scaling curve and the scale parameters $t_0,\ h_0,\ T_c$ could be extracted. 
As the calculations had been performed with staggered fermions the scaling
analysis has been performed by comparing results with the magnetic equation
of state for a $O(2)$ universality class rather than $O(4)$ as one should 
find in the continuum limit for two massless quark flavor. However, as has 
been argued in \cite{magnetic} both scaling curves are similar in the limited
range of z values, where this scaling analysis has been performed, and
the scaling analysis could have been performed with $O(4)$ scaling functions
as well.

At least on these coarse $N_\tau=4$ 
lattices violations of scaling have been found to be small also for physical
values of the light quark mass, {\it i.e.} $m_l/m_s\simeq 1/27$.
Of course, it is to be expected that the scale parameters, extracted on
coarse lattices with temporal extent $N_\tau=4$, are subject to cut-off effects.
We therefore extend the analysis of Ref.~\cite{magnetic} to smaller lattice
spacings. We perform calculations on lattices with temporal extent $N_\tau =8$. 
We follow here the discussion presented in Ref.~\cite{magnetic} 
and introduce two order parameters, that are multiplicatively renormalized
by multiplying the chiral condensate with the strange quark mass, but differ 
in handling additive divergences, linear in the quark mass\footnote{ At finite values 
of the cut-off these  terms are, of course, finite and may be viewed as a specific
contribution to the regular part that will not alter the scaling properties for 
sufficiently small values of the quark mass.},
\begin{eqnarray}
M_b &\equiv& N_\tau^4 \hm_s \langle \bar{\psi}\psi \rangle_l \; , \nonumber \\
M &\equiv&  N_\tau^4 \hm_s \left( \langle \bar\psi \psi \rangle_l -
\frac{m_l}{m_s} \langle \bar\psi \psi \rangle_s \right) \; .
\label{order_parameter}
\end{eqnarray}
We use in our study data for the chiral condensate which have been
collected by the hotQCD \cite{hotQCD} and RBC-Bielefeld \cite{eos}
collaborations in their studies of the (2+1)-flavor equation of state
on lattices of size $32^3\times 8$ as well as in the course
of analyzing the transition temperature in (2+1)-flavor QCD \cite{hotQCDTc}. 
These calculations have been performed with the p4-action \cite{p4action}
for three different quark mass 
ratios $m_l/m_s=0.2, \ 0.1$ and $0.05$, respectively. 
We use subsets of these data samples which cover a small temperature
interval close to the transition region, but covering also the region
of the transition temperature in the chiral limit. For the smallest 
quark mass ratio this includes a set of
16 values of the gauge coupling which cover a 
temperature interval $0.95\lsim T/T_c \lsim 1.12$ ($3.48< 6/g^2 < 3.545$). 
Typically (20.000-30.000) trajectories of 0.5 time-units have been generated 
for each set of quark masses and gauge couplings. 
The temperature scale used in our scaling analysis is based
on calculations of the scale parameter $r_0$ that has been determined from
calculations of the heavy quark potential performed on lattices of size $32^4$ 
\cite{hotQCD,eos}.

The basic approach for the scaling analysis on $N_\tau =8$ lattices
is identical to that described in Ref.~\cite{magnetic}. However,
as the calculations on the $N_\tau =8$ lattices have not been performed for
as small light quark masses as in the $N_\tau=4$ analysis \cite{magnetic}, 
where the smallest ratio was $m_l/m_s=1/80$, we did not perform a separate
analysis for the determination of scaling parameters in the small mass
regime and a determination of scaling violating terms for larger values
of the quark masses, as it has been done in Ref.~\cite{magnetic}.
We perform a simultaneous analysis of data obtained for all three quark mass 
ratios and include scaling violating regular terms in the ansatz for scaling 
fits,
\begin{equation}
M(t,h) = h^{1/\delta} f_G(t/h^{1/\beta\delta}) + a_t \Delta T H + b_1 H + b_3 H^3\; .
\label{violations}
\end{equation}
with $\Delta T = (T-T_c)/T_c$ and $H=m_l/m_s$. 
This includes all the scaling violating
terms also used on $N_\tau =4$ lattices. In our final analysis, however, 
we will set $b_3=0$.

\begin{figure}[t]
\begin{center}
\epsfig{file=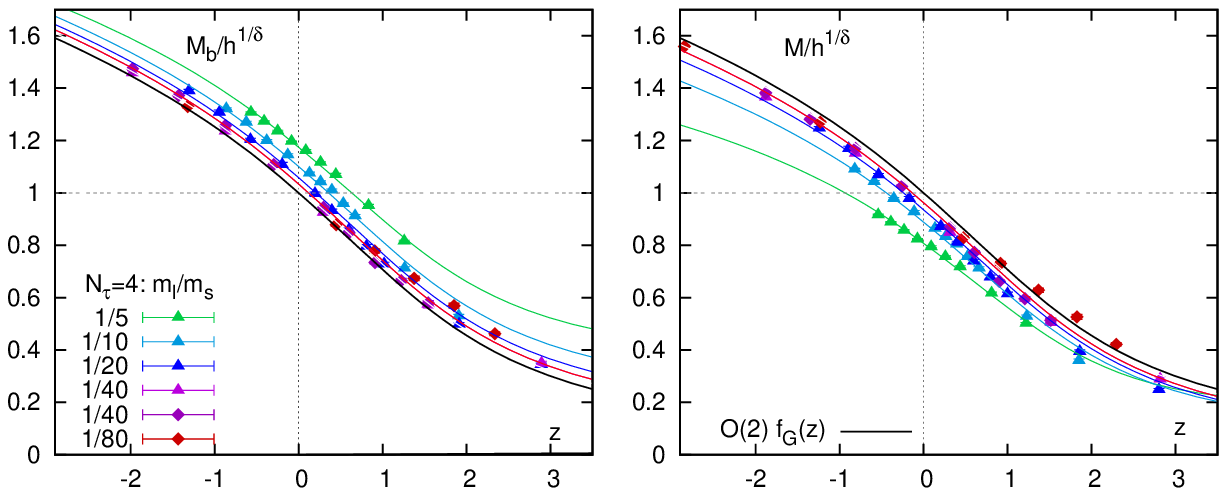,width=15.0cm}
\epsfig{file=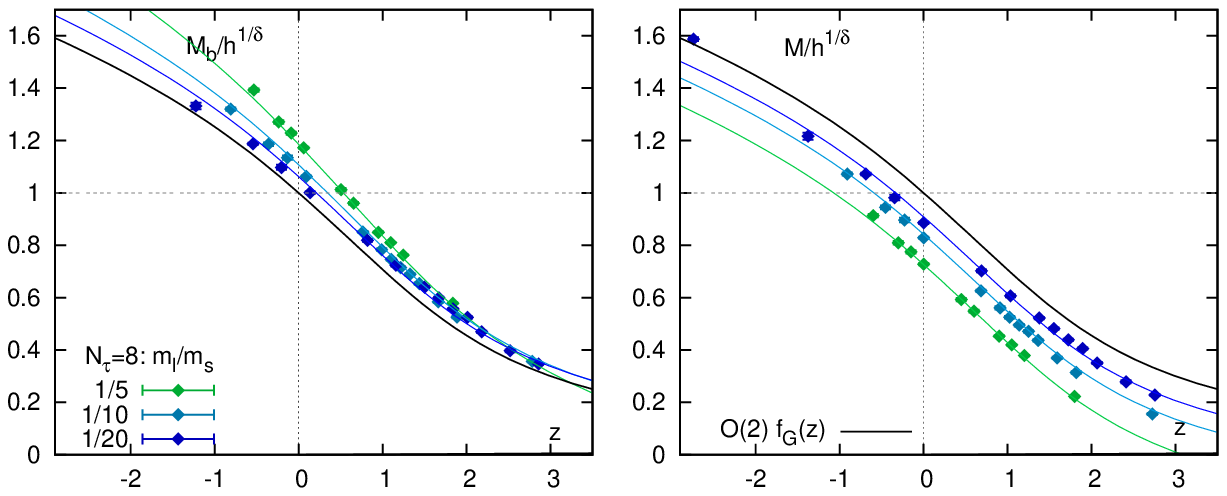,width=15.0cm}
\end{center}
\caption{\label{fig:scalingNt8}
Fit of the $O(2)$ scaling function to
numerical results for the subtracted order parameter $M$ (right) and the
non-subtracted light quark condensate $M_b$ (left), both for $N_\tau=4$ (top) 
and $N_\tau=8$ (bottom). The fits include an
ansatz for violations of scaling as discussed in the text. Shown are results 
for $m_l/m_s\le 1/5$.
}
\end{figure}

We verified that this approach, applied to the $N_\tau=4$ data set, and
restricted to the same mass range available now on $N_\tau=8$ lattices,
{\it i.e.}  $1/20\le m_l/m_s\le 1/5$,
leads to results compatible with our earlier findings.
We  extract values for $t_0$, $h_0$ and $T_c$ which are
similar to those obtained previously. In fact, results for fit 
parameters obtained from the analysis of different order 
parameters, $M$ and $M_b$, turn out to be in even better agreement. 
Results of this new scaling analysis for the $N_\tau =4$ data set
is shown in the upper half of Fig.~\ref{fig:scalingNt8}. All fit 
parameter are summarized in Table~\ref{tab:parameter}.

Using the approach described above for the analysis of our $N_\tau=8$ data 
set we find good agreement of fit parameters extracted from an analysis of 
$M_b$ and $M$, respectively. Results of this scaling analysis are shown in 
the lower half of Fig.~\ref{fig:scalingNt8}. All fit
parameter are summarized in Table~\ref{tab:parameter}.
As already noted in the analysis performed on lattices with temporal extent
$N_\tau=4$ \cite{magnetic}, we observe also for $N_\tau=8$ 
that scaling violations are small for $m_l/m_s\leq 1/10$. This confirms that 
physical quark mass values, corresponding to $m_l/m_s \simeq 1/27$, are in the scaling 
region. 

The constants $h_0$, $t_0$ determined for $N_\tau=8$ take on values different 
from those for $N_\tau=4$. The invariant combination of 
scale parameters, 
\begin{equation}
z_0\equiv h_0^{1/\beta\delta}/ t_0=z_0(m_s)+\mathcal{O}(a^2) \; ,
\end{equation}
changes by about 50\% which shows that its continuum extrapolation is
not yet possible. This also is the case for $t_0$ and $h_0$ separately.

\begin{table}
\begin{center}
\vspace{0.3cm}
\begin{tabular}{|c|c|c|c|c|c|c|c|}
\hline
 $N_\tau$ & $M_i$ & $t_0$ & $h_0$ & $T_c(0)$~[MeV] & $a_t$ & $b_1$ & $z_0$ \\
\hline
\multicolumn{8}{|c|}{fit using the scaling term only}\\
\hline
4 & $M_b$ & 0.0037(2) & 0.0022(3) & 194.5(4) & - & - & 6.8(5) \\
  & $M$   & 0.0048(5) & 0.0048(2) & 195.6(4) & - & - & 8.5(8) \\
\hline
\multicolumn{8}{|c|}{fit using scaling and regular terms}\\
\hline
4 & $M_b$ & 0.00407(9) & 0.00295(22) & 194.9(2) & 3.8(21)  &  2.1(1) & 7.5(3) \\
  & $M$   & 0.00401(9) & 0.00271(20) & 194.8(2) & 11.2(21) & -2.4(1) & 7.2(3) \\
8 & $M_b$ & 0.00271(21) & 0.00048(9) & 174.1(8) & -10.1(16) & 3.3(5) & 3.8(5) \\
  & $M$   & 0.00302(22) & 0.00059(10) & 175.1(8) & -0.9(15) & -4.6(4) & 3.8(4) \\
\hline
\end{tabular}
\end{center}
\caption{Scale parameters determined from the scaling fits on lattices
of temporal extent $N_\tau=4$ and $8$. In column 6 and 7 we list the
couplings for the leading scaling violating corrections. The last column 
gives $z_0\equiv h_0^{1/\beta\delta}/ t_0$. 
We give the results for parameters entering the definition of scaling 
functions for $M_b$ and the subtracted order parameter $M$ as defined 
in Eq. \ref{order_parameter}. Only the former has been used in the analysis 
of the mixed susceptibilities. Note that fits including regular terms, give 
consistent determinations of the parameters of the scaling functions 
determined from $M_b$ and $M$, respectively.
}
\label{tab:parameter}
\end{table}

When comparing results obtained for $N_\tau=4$ and $N_\tau=8$ 
one also has to take into account the dependence of the scale parameters on 
the strange quark mass. In fact, as the scaling analysis has been performed at 
bare strange quark mass values $m_s$, fixed in lattice units, the 
corresponding physical value in the chiral limit at $t=0$ is only determined 
a posteriori, once $T_c$ has been determined. It turns out that the physical 
values of the strange quark mass in the $N_\tau=4$ and $8$ calculations differ 
at $T_c$ by about 10\%. One may account for this mismatch by reweighting the 
results for the chiral condensates in the light and strange quark masses 
\cite{unger}. However, we will not attempt to do this here.

The main outcome of the $N_\tau=8$ scaling analysis, aside from confirming 
the good scaling properties of the order parameter at a twice smaller value  
of the lattice spacing, is a determination of the scale parameters 
and the transition temperature $T_c$ in the chiral limit, needed in 
the definition
of the scaling variable $z$, {\it i.e.} the determination of $t_0,\ h_0$ and
$T_c$. We summarize these results in Table~\ref{tab:parameter}.
In the next section we will make use of these scale parameters to determine 
the curvature of the phase transition line for small values
of the quark chemical potential. 

\section{Curvature of the critical line}
\label{sec:curve}
As outlined in the beginning of the previous section at leading order
the light quark chemical potential only enters the reduced 
temperature $t$, as introduced in Eq.~\ref{scalingfields}. 
Also at non-vanishing values of the quark chemical potential 
the phase transition point is located at $t=0$. 
The variation of the transition temperature with chemical potential
therefore is parametrized in terms of the constant
$\kappa_q$ introduced in Eq.~\ref{scalingfields},
\begin{equation}
\frac{T_c(\mu_q)}{T_c} = 1 -\kappa_q \left( \frac{\mu_q}{T} \right)^2 
+{\cal O}\left(\left( \frac{\mu_q}{T}\right)^4\right) \; .
\label{criticalline}
\end{equation}
To determine the chiral phase transition line in the $T$-$\mu$ plane we thus
need to determine the proportionality constant $\kappa_q$. 
This is, in fact, 
the only left over free parameter in universal scaling functions that needs to 
be determined. 
All other parameters 
($t_0,\ h_0,\ T_c\equiv T_c(\mu_q=0)$) 
have already been determined in the scaling analysis of the order parameter 
discussed in the previous section.

The constant $\kappa_q$  can be determined by analyzing the 
dependence of the chiral condensate on the light quark chemical potential.
Of course, at vanishing light quark mass one would simply determine the
temperature at which $\langle \bar{\psi}\psi\rangle_l$ vanishes. At 
non-zero but small values of the quark mass this information is encoded
in scaling functions. To extract information about the dependence of 
the scaling variable $t$ on $\kappa_q$ it suffices to consider the 
leading order Taylor expansion coefficient of the chiral condensate,
\begin{equation}
\frac{\langle \bar{\psi}\psi\rangle_l}{T^3} =  
\left( \frac{\langle \bar{\psi}\psi\rangle_l}{T^3} \right)_{\mu_q=0} +
\frac{\chi_{m,q}}{2T} \left(\frac{\mu_q}{T}\right)^2 + 
{\cal O}((\mu_q/T)^4) \; ,
\label{pbp_Taylor}
\end{equation}
where 
\begin{equation}
 \frac{\chi_{m,q}}{T} =  \frac{\partial^2 \langle 
\bar{\psi}\psi\rangle_l/T^3}{\partial (\mu_q/T)^2}
= \frac{\partial \chi_q/T^2}{\partial m_l/T}  \; .
\label{mixed}
\end{equation}
The mixed susceptibility $ \chi_{m,q}$ is proportional to the leading 
order coefficient of the Taylor expansion of the chiral condensate, which 
has been introduced in \cite{Ray,Taylor6}. It 
may also be viewed as the quark mass derivative of the light quark number 
susceptibility ($\chi_q$). Details of its
definition in terms of inverses of the staggered fermion matrix and its
derivatives with respect to the quark chemical potential 
are given in Appendix A of Ref.~\cite{Taylor6}. We have summarized the 
formulas relevant for our current analysis in an Appendix. 

In the massless limit the chiral order parameter vanishes at $T_c$ and 
varies as $M\sim (-t)^{\beta}$.
Its derivative with respect to $t$ thus will diverge at 
$T_c$ like $t^{\beta-1}$. The same singular behavior will thus show up
in a derivative of the chiral condensate with respect to temperature as 
well as the second derivative with respect to $\mu_q/T$. The 
pre-factors of the singularity in ${\rm d} M/{\rm d}T$ and 
${\rm d}^2 M/{\rm d}(\mu_q/T)^2$, however, will differ by a 
factor 
$2\kappa_q T_c$. We will make use of this relation to determine the 
curvature of the critical line at $\mu_q=0$.

In the vicinity of the critical point the mixed susceptibility can be 
expressed in terms of the scaling function 
$f'_G(z)\equiv {\rm d}f_G(z)/{\rm d} z$,  
\begin{equation}
\frac{\chi_{m,q}}{T} =  
\frac{2 \kappa_q T}{t_0 m_s} h^{-(1-\beta)/\beta\delta}  f'_G(z) \;  .
\label{mixedscaling}
\end{equation}
The scaling function $f'_G(z)$ is easily obtained from $f_G(z)$
by using the implicit parametrization for the latter given in 
Ref.~\cite{engelsO2}. We also note that $ \chi_{m,q}$ diverges as 
function of the light quark mass at $t=0$, {\it i.e.} at the chiral 
phase transition temperature. In contrast to
the chiral susceptibility, $\chi_m \sim \partial M/ \partial m_l$, which 
stays finite in the chiral limit only for $t>0$, the mixed susceptibility 
is finite for all $t\ne 0$; for $t<0$ it scales like $(-t)^{\beta-1}$ 
while for $t>0$ it behaves like $h t^{-1-\gamma}$.

For small values of the light quark mass numerical results for the 
mixed susceptibilities
$ \chi_{m,q}$ may be compared to the right hand side of 
Eq.~\ref{mixedscaling}. Here all parameters that enter $f'_G(z)$ are
known and the only  undetermined parameter is $\kappa_q$. 
As we did for the analysis of the magnetic equation of
state, we should consider the influence of scaling violations induced by 
non-zero values of the light quark masses on the determination of the 
curvature of the phase transition line. The leading quark mass corrections 
identified in the analysis of the magnetic equation of state will not
contribute to $ \chi_{m,q}$ 
as they do not depend on the chemical potential. 
The first scaling violating term would 
arise from a regular term that gives corrections to the order parameter of the 
form $M \sim a_q H (\mu_q/T)^2$. This would give rise to 
corrections to the scaling relation given in Eq.~\ref{mixedscaling}
\begin{equation}
t_0 h^{(1-\beta)/\beta\delta} \frac{m_s}{T} \frac{\chi_{m,q}}{T} =
2 \kappa_q   f'_G(z)  
+ \frac{2 a_q h_0^{1/\delta} }{z_0}  H^{1+(1-\beta)/\beta\delta}   
\; .
\label{mixedscaling2}
\end{equation}
As we do not know the pre-factor $a_q$ we need to check
in the analysis of the mixed susceptibilities whether corrections to 
scaling play a role. We note, however, that in the case of the 
magnetic equation of state the dominant corrections arise from the 
term $b_1 H$, which is ultraviolet divergent in the continuum limit.
This term drops out in the analysis of $\chi_{m,q}$.
Moreover, scaling violating terms are further suppressed by a factor
$H^{(1-\beta)/\beta\delta}\equiv H^{0.39}$ (or $H^{0.34}$ in $O(4)$
symmetric models) as the dominant scaling term itself is divergent 
at $(t,h)=(0,0)$. We thus expect scaling violations to be small.

Using a subset of the data samples described in the previous section, 
we calculated the mixed susceptibility $ \chi_{m,q}$ on lattices with 
temporal extent 
$N_\tau=4$ for several values of the quark mass. For this analysis we used 
data sets separated by 50 trajectories. For the lightest quark mass ratio,
$m_l/m_s=1/80$, we selected 4 and for the three heavier quark mass ratios,
$m_l/m_s=1/10,\ 1/20, \ 1/40$, we choose 6 values of the gauge coupling in a 
narrow temperature interval close to the chiral phase transition temperature 
$T_c$, {\it i.e.} $-0.02\le (T-T_c)/T_c \le 0.06$.
Typically this involved about 500 
to 950 gauge field configurations per parameter set, except for 
the lightest quark mass ratio where we analyzed about 350 
gauge field configuration.
On each gauge field configuration we calculated the 
various operators necessary to construct $ \chi_{m,q}$. 
Explicit expression for them are given in the Appendix (see also 
Appendix A of \cite{Taylor6}).

The calculation of the various operators required inversions of the 
staggered fermion matrix with a large set of random noise vectors. We
used 500  noise vectors on each gauge field configuration and constructed 
unbiased estimators for the various traces that need to be calculated. 
All these calculations could be performed very efficiently on a GPU cluster.
 
\begin{figure}[t]
\begin{center}
\epsfig{file=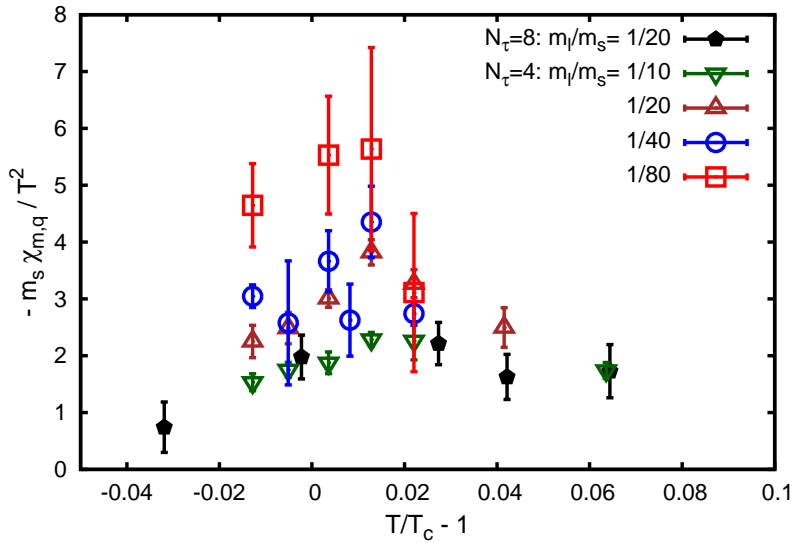,width=11.0cm}
\end{center}
\caption{\label{fig:c2pbpiq}
The mixed light quark number susceptibility 
as a function of the reduced temperature, $(T-T_c)/T_c$.
Shown are results obtained at two values of the cut-off, $N_\tau =4$
(open symbols) and $N_\tau =8$ (filled symbols), and for several values of 
the light to strange quark mass ratio. 
}
\end{figure}

Results obtained for the mixed light quark number susceptibility,
$ \chi_{m,q}$, on lattices with temporal
extent $N_\tau=4$ are shown in Fig.~\ref{fig:c2pbpiq}.
We clearly see that $ \chi_{m,q}$ 
increases in the transition region with decreasing values 
of $m_l/m_s$. 

\begin{figure}[t]
\begin{center}
\epsfig{file=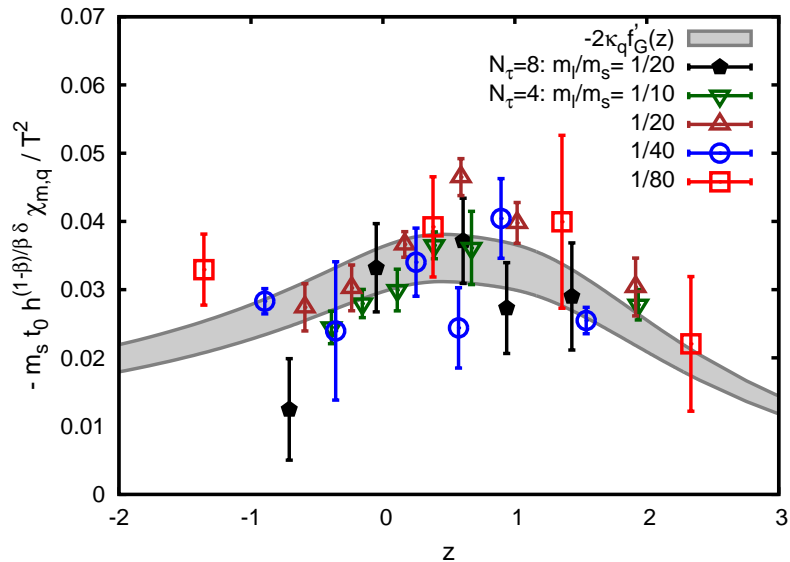,width=11.0cm}
\end{center}
\caption{\label{fig:c2pbp}
The scaled mixed susceptibility as function of the scaling variable
$z=t/h^{1/\beta\delta}$. The data are compared to the $O(2)$ 
scaling curve. The band shows a 10\% error band on this curve which
arises from statistical errors on the calculated observables as well as 
from the errors on the scaling parameters $t_0$ and 
$z_0$ given in Table~\ref{tab:parameter}.
}
\end{figure}

Using the scaling relation given in 
Eq.~\ref{mixedscaling} we can re-scale the data and obtain a unique scaling
curve. This scaling curve can be mapped onto the $O(2)$ scaling function 
$f'_G(z)$ with a simple multiplicative rescaling factor, $2\kappa_q$.
The resulting scaling plot is shown in Fig.~\ref{fig:c2pbp}.
To check for possible contributions from scaling violating terms we have 
analyzed the data separately for quark mass ratios $m_l/m_s = 1/10,\ 1/20$ 
and $m_l/m_s = 1/40,\ 1/80$. These fits agree within statistical errors.
We then determine the curvatures $\kappa_q$ from fits 
to the complete data set. Results of these fits are summarized in 
Table~\ref{tab:fit}.
\begin{table}[t]
\begin{center}
\vspace{0.3cm}
\begin{tabular}{|c|c|c|c|}
\hline
 $N_\tau$ & $m_l/m_s$ & $\kappa_q$ & $\chi^2$/dof \\
\hline
4  & 1/10, 1/20 & 0.0598(26) & 3.5 \\
~  & 1/40, 1/80 & 0.0573(29) & 1.5 \\
\hline
8  & 1/20 & 0.0559(35) & 0.4 \\
\hline
4,\ 8  & all & 0.0591(17) & 2.1 \\
\hline
\end{tabular}
\end{center}
\caption{Determination of the curvature of the critical surface
of the chiral phase transition in $(2+1)$-flavor QCD as function of 
the light quark chemical potential
$\mu_q$. The table summarizes fits performed separately
for two lighter and two heavier quark mass sets as well as the combined
data set.
}
\label{tab:fit}
\end{table}

The scaling analysis performed for the mixed susceptibility on lattices with 
temporal extent $N_\tau =4$ suggests that the determination
of the curvature parameter $\kappa_\mu$ can be reliably performed with 
quark masses $m_l/m_s\lsim 1/10$. This is in accordance with the scaling
analysis of the order parameter itself, which we have discussed in the 
previous section. It thus seems to be safe to extract the curvature parameter
also at smaller values of the lattice spacing, {\it i.e.} from our $N_\tau =8$
data set, by using the smallest quark mass ratio available there, 
$m_l/m_s=1/20$.
We have performed calculations at five values of the temperature using gauge 
field configurations on $32^3\times 8$ lattices generated by the hotQCD
collaboration \cite{hotQCDTc}. 
For these parameter sets we have analyzed 300 to 600 gauge field 
configurations, 
which were separated by 100 trajectories. Again we used 500 noise vectors
for the calculation of all relevant operators on each of the gauge field
configurations.
The result of this analysis is shown in Fig.~\ref{fig:c2pbpiq} and
Fig.~\ref{fig:c2pbp} with filled symbols. As can be seen they agree well 
with results obtained on coarser lattices. 

When rescaling data obtained for $ \chi_{m,q}$ to the 
$O(2)$ scaling curve $f'_G(z)$ we need to take into account errors on
the scaling parameters $t_0$ and $z_0$ (or $h_0$). This leads to a 10\% 
error for the determination of the curvature terms.

Performing a combined fit to all results obtained for different quark 
mass values and lattice spacings we obtain
\begin{equation}
\kappa_q = 0.059 (2)(4) \;  .
\label{kappa}
\end{equation}

This result for the curvature of the critical line is about a factor two 
larger than the reweighting results obtained in (2+1)-flavor QCD 
\cite{Fodor}. It is however consistent with results obtained in 
calculations with imaginary chemical potentials. In fact it lies inbetween
the 2-flavor \cite{analytic} and 3-flavor \cite{deForcrand3} simulations
performed with the standard staggered fermion formulation and also is
consistent with results reported from (2+1)-flavor simulations with 
imaginary chemical potential performed with
the action used also in this study (p4-action) \cite{lat10}.

\section{Conclusions}

With this analysis we have established a systematic way to determine the
curvature of the QCD phase transition line in the chiral limit for
small values of the light quark chemical potential. We have determined the
curvature for two values of the cutoff using an improved staggered fermion
action (p4-action). Within our present statistical accuracy we did not 
observe any significant quark mass dependence of the scaled mixed 
susceptibilities. The result observed for the curvature of the second
order phase transition line in the chiral limit thus also is a good
estimate for the crossover line at physical values of the light quark 
masses.

Although the final result for the curvature term seems to show little
cutoff dependence, one has to be cautious as the other three  scale 
parameters that enter this analysis ($t_0$, $h_0$ and $T_c$) all vary
significantly as the lattice spacing is reduced by a factor of two. 
Clearly more work is needed to extrapolate safely to the continuum limit
We will do so in the future by repeating the analysis with a discretization
scheme that also suppresses cut-off effects arising from taste symmetry
violation in the staggered fermion action more efficiently (hisq action).

In our current analysis we have kept the strange quark and 
isospin chemical potentials equal to zero. A direct comparison to 
the situation met in heavy ion collisions on the freeze-out curve
thus should be done with caution. 
However, at least for large energies, 
{\it i.e.}, small values of the baryon chemical potential 
experimental results for the freeze-out curve correspond to electric
charge chemical potentials, which are more than an order of magnitude
smaller than $\mu_B$. As susceptibilities obtained by derivatives with
respect to strange quark chemical potentials rather than light quark chemical
potentials are generally smaller, one may also expect that the curvature in 
the $\mu_S$ direction will turn out to be smaller. 
It thus seems that the curvature of the critical
surface along the $\mu_q\simeq \mu_B/3$ direction is most relevant
for a comparison of lattice QCD results with the experimentally 
determined freeze-out curve. 
The phenomenological parametrization of the freeze-out curve given in
\cite{cleymans} yields 
$T_{freeze}(\mu_B)/T_{freeze}(0) \simeq 1 -0.21(2) (\mu_q/T)^2+{\cal O}(\mu_q^4)$.
The curvature of the freeze-out curve thus is about 
a factor 4 larger than that determined here for the chiral phase 
transition curve. 
This suggests that the freeze-out curve may not follow
the chiral phase transition or crossover line at non-zero values of the 
chemical potential. With increasing $\mu_q/T$ the hadronic freeze-out
seems to happen further away from criticality.
At the largest value of the light quark chemical
potential currently explored in the low energy scan at RHIC
\cite{STAR}, $\mu_q/T \simeq 1$,
the freeze-out temperature may be about 15\% below the crossover temperature. 

Nonetheless, as pointed out above, one still needs to 
improve the current lattice calculations. Results closer to the 
continuum limit with further improved fermion discretization schemes
are needed and one should also get control over the influence of
non-vanishing strange quark chemical potentials in order
to firmly establish the separation of the freeze-out curve from the chiral
transition line as advocated above.  
 
\section*{Acknowledgments}
\label{ackn}
This work has been supported in part by contracts DE-AC02-98CH10886
with the U.S. Department of Energy, the BMBF under grant 06BI401, the 
Gesellschaft f\"ur Schwerionenforschung under grant BILAER, the Extreme 
Matter Institute under grant HA216/EMMI and the Deutsche
Forschungsgemeinschaft under grant GRK 881. 
CS has been partially supported through the Helmholtz International Center
for FAIR which is part of the Hessian LOEWE initiative.
Numerical simulations have been performed on the BlueGene/L at the New York 
Center for Computational Sciences (NYCCS) which is supported by the U.S. 
Department of Energy and by the State of New York, the GPU cluster of USQCD 
at Jefferson Laboratory, the GPU cluster SCOUT at the Center for Scientific 
Computing (CSC) at Frankfurt University, 
as well as the John von Neumann Supercomputer center (NIC) at 
FZ-J\"ulich, Germany. We thank M. Bach for his help in developing the
CUDA based programs used for our data analysis on GPU clusters.

\appendix
\section{\boldmath The mixed susceptibility 
$ \chi_{m,q}$}
We summarize here the operators entering a calculation of the 
mixed susceptibility $ \chi_{m,q}$ introduced in Eq.~\ref{mixed}.
This susceptibility is proportional to the second order Taylor expansion 
coefficient of the chiral condensate in terms of the light quark 
chemical potentials. 
Using Appendix A of Ref.~\cite{Taylor6}) we introduce
the expectation value of the light quark chiral condensate as
\begin{equation}
\frac{\left\langle \bar{\psi} \psi \right\rangle_l}{T^3} = \langle {\cal C}_0
\rangle \; ,
\label{pbpa}
\end{equation}
and obtain for the mixed susceptibility,
\begin{eqnarray}
\frac{\chi_{m,q}}{T}  &\equiv& 
\frac{\partial^2 \left\langle \bar{\psi} \psi \right\rangle_l /T^3
}{\partial (\mu_{q}/T)^2}  \biggr|_{\mu_q=0} =
\frac{1}{N_{\tau}^2} \frac{\partial^2 
\left\langle \bar{\psi} \psi \right\rangle_l/T^3}{\partial \hmu^2} 
\biggr|_{\hmu=0}
\nonumber \\
&=& \frac{1}{N_{\sigma}^3 } 
\left( 
\left\langle {\cal C}_2 \right\rangle 
+ 2 \left\langle {\cal C}_1 {\cal D}_1 \right\rangle 
+ \left\langle {\cal C}_0 {\cal D}_2 \right\rangle 
+ \left\langle {\cal C}_0 {\cal D}_1^2 \right\rangle
 - 
\left\langle {\cal C}_0 \right\rangle 
\left( 
\left\langle {\cal D}_2 \right\rangle 
+\left\langle {\cal D}_1^2 \right\rangle 
\right)
\right) \; , 
\end{eqnarray}
where we have introduced the shorthand notation $\hmu=\mu_qa$ for
the chemical potential expressed in units of the lattice spacing.
Here ${\cal C}_n$ and ${\cal D}_n$ denote n-th derivatives of the 
trace of the inverse fermion matrix ($D$) and logarithms of its
determinant, respectively,
\begin{eqnarray}
{\cal C}_n = \frac{1}{4} \frac{\partial^n {\rm tr} D^{-1}} 
{\partial \hmu^n}
\quad , \quad 
{\cal D}_n = \frac{1}{2} \frac{\partial^n \ln \det D} 
{\partial \hmu^n} \quad , 
\label{eq:basic}
\end{eqnarray}
where the derivatives are defined with respect to the flavor
chemical potential $\hmu\equiv\hmu_f$ for the quark flavor $f$. 
The factors $1/4$ and $1/2$ arise because we define the mixed 
susceptibility as a derivative of the 1-flavor light quark chiral 
condensate with respect to the light quark chemical potential $\mu_q$
which is identical for the two light flavor components of the fermion
action, {\it i.e.} a flavor factor $n_f=2$ arises only in derivatives
of the logarithm of the fermion matrix. 

The operators needed to calculate $\chi_{m,q}$ thus are
\begin{eqnarray}
{\cal C}_0
&=& \frac{1}{4}  
 {\rm tr} \left( D^{-1} \right) \\
{\cal C}_1
&=& - \frac{1}{4}  
 {\rm tr} \left( D^{-1} \frac{\partial D}{\partial \hmu}
 D^{-1} \right) \\
{\cal C}_2
&=& -\frac{1}{4}  \left(
{\rm tr} \left( D^{-1} \frac{\partial^2 D}{\partial \hmu^2}
 D^{-1} \right)
 - 2 {\rm tr} \left( D^{-1} \frac{\partial D}{\partial \hmu}
    D^{-1} \frac{\partial D}{\partial \hmu} D^{-1} \right) 
\right)  \\
{\cal D}_1
&=&  \frac{1}{2}
{\rm tr} \left( D^{-1} \frac{\partial D}{\partial \hmu} \right) ,
\label{eq:dermu1} \\
{\cal D}_2
&=&  \frac{1}{2} \left(
{\rm tr} \left( D^{-1} \frac{\partial^2 D}{\partial \hmu^2} \right)
 - {\rm tr} \left( D^{-1} \frac{\partial D}{\partial \hmu}
                   D^{-1} \frac{\partial D}{\partial \hmu} \right)
\right) ,
\label{eq:dermu2} 
\end{eqnarray}

 
\end{document}